\documentclass[prd,superscriptaddress,reprint,showpacs,preprintnumbers,amsmath,amssymb]{revtex4-1}
\usepackage{graphicx}
\usepackage{bm}
\usepackage[breaklinks, colorlinks=true,  linkcolor=red, citecolor=blue, urlcolor=blue]{hyperref}

\def\delsla{\!\not\!\partial}

\begin{document}

\title{Impact of vector-current interactions on the QCD phase diagram}

\author{Thomas Hell}
\affiliation{Physik-Department, Technische Universit\"{a}t M\"{u}nchen, D-85747 Garching, Germany}
\affiliation{ECT*,
Strada delle Tabarelle 286,
I-38123 Villazzano (Trento),
Italy}

\author{Kouji Kashiwa}
\affiliation{RIKEN/BNL Research Center, Brookhaven National Laboratory,
             Upton, NY-11973, USA}

\author{Wolfram Weise}
\affiliation{Physik-Department, Technische Universit\"{a}t M\"{u}nchen, D-85747 Garching, Germany}
\affiliation{ECT*,
Strada delle Tabarelle 286,
I-38123 Villazzano (Trento),
Italy}


\begin{abstract}
Using a nonlocal version of the Polyakov-loop-extended Nambu--Jona-Lasinio model, we investigate effects of a nonderivative vector-current interaction (relating to the quark-number density) at both real and imaginary chemical potentials.
This repulsive vector interaction between quarks has the following impact on the chiral first-order phase transition: at imaginary chemical potential it sharpens the transition at the Roberge-Weiss (RW) end point and  moves this critical point toward lower temperatures; at real chemical potential, the critical end point moves on a trajectory towards larger chemical potentials and lower temperatures with increasing vector coupling strength. The conditions are discussed at which the first-order phase transition disappears and turns into a smooth crossover.
\end{abstract}

\maketitle

\section{Introduction}
Exploring the phase diagram of quantum chromodynamics (QCD)  at finite temperature and real chemical potential is one of the most interesting and important subjects in particle and nuclear physics. Lattice-QCD (LQCD) simulations are a powerful method to investigate the QCD thermodynamics at zero chemical potential.
At finite chemical potential, however, LQCD suffers from the so-called sign problem which restricts the applicability of LQCD to the region of small real chemical potential ($\mu_{\rm R}$) and high temperatures (see Ref.~\cite{Forcrand:2009}).
Therefore, model calculations (admittedly with substantial ambiguities \cite{Kashiwa:2007}) are used to investigate the phase structure at moderate and large $\mu_\mathrm{R}$.

A promising strategy for studying the QCD phase diagram at finite $\mu_\mathrm{R}$ is the imaginary-chemical-potential matching approach~\cite{Kashiwa:2009yd}.
It is similar to the usual imaginary chemical potential approach for LQCD
~\cite{Forcrand:2002ls,*Forcrand:2003gd,
       D'Elia:2003di,*D'Elia:2004ns,Chen:2005sa,Wu:2007iu}:
LQCD data at finite imaginary chemical potential ($\mu_\mathrm{I}$) are extrapolated to the $\mu_\mathrm{R}$ region by using an analytic function.
In the imaginary-chemical-potential matching approach we extract some important restrictions for the model design from the $\mu_\mathrm{I}$ region. This allows us to extend the model to the real-chemical-potential region more realistically (cf.\ Ref.~\cite{Kashiwa:2011}). 
The important point is that the $\mu_\mathrm{I}$ region  encodes almost all information of the $\mu_\mathrm{R}$ region.  
This fact can be understood through a Fourier transformation of the grand-canonical partition function, $\mathcal{Z}$, in terms of $\theta = \mu_\mathrm{I}/T$ in the case that the baryon number is a good quantum number:
\begin{align}
\mathcal{Z}_\mathrm{C} (N_q) 
&= \frac{1}{2\pi} \int^\pi_{-\pi} \mathrm{d} \theta\, 
   e^{-iN_q \theta} \mathcal{Z} (\theta)\,.
\end{align}
Here, $\mathcal{Z}_\mathrm{C}(N_q)$ is the \emph{canonical} partition function with real quark numbers ($N_q$).
Moreover, QCD possesses the so-called Roberge-Weiss (RW) periodicity~\cite{Roberge:1986}:  thermodynamical quantities have a periodicity of $2\pi/3$ along the $\theta$-axis.
This periodicity is described by invariance under the extended ${\mathbb
Z}_3$ symmetry~\cite{Sakai:2008}
\begin{equation}
\theta \to \theta + 2\pi k/3\,,~~~~
\Phi \to e^{-2\pi k/3}\Phi\,,~~~~
{\bar \Phi} \to e^{2\pi k/3}{\bar \Phi}\,,
\end{equation}
with integer $k$, where $\Phi$ is the Polyakov loop and $\bar \Phi$ its conjugate.
The RW periodicity enables us to determine which interactions are relevant and how strong the couplings are by comparing model results with LQCD data at finite $\mu_\mathrm{I}$.
Note, that $\mu_\mathrm{I}$ can be absorbed in the boundary angle of the temporal direction of the quark field.
From this viewpoint, quarks are fermions at $\theta = 2\pi k/N_c$ and these become  boson-like at $\theta = \pi (2 k + 1)/N_c$.
Therefore, the dual quark condensate was proposed
 \cite{Bilgici:2008mr} as an order parameter for the chiral and the deconfinement phase transition.

At $\theta=\pi/3$ another characteristic property of QCD, the
so-called RW transition, arises: this RW transition can be related to charge-conjugation (${\cal C}$) or ${\cal C} {\mathbb Z}_3$ symmetry breaking
~\cite{Kashiwa:2012xm}. ${\cal C}{\mathbb Z}_3$ symmetry is explicitly broken at finite
$\mu_\mathrm{R}$, but it is not explicitly broken at
$\theta=\pi/3$ because of  RW periodicity (see, e.\,g., Ref.~\cite{Kouno:2009}).
On the RW transition line, $\theta$-odd quantities can have a finite
value, but they vanish for temperatures below the RW end point \cite{Roberge:1986}.
Therefore, we can interpret $\theta$-odd quantities as 
order parameters of spontaneous ${\cal C}$-symmetry breaking.

The Polyakov-loop-extended Nambu--Jona-Lasinio (PNJL) model is a promising approach as it preserves the RW periodicity in the same way as QCD.
In the present study we extend the nonlocal version of the two-flavor PNJL model from Refs.~\cite{Hell:2009,Hell:2011} by introducing a  nonderivative vector-current interaction between quarks both at imaginary and real chemical potentials.

This paper is organized as follows: 
In section \ref{frameworksection} we introduce the nonlocal PNJL model that is used in our calculations. In particular, we describe in detail the treatment of the vector-type interaction in the nonlocal framework. We show how this approach can be extended to imaginary chemical potentials.  
Section \ref{resultssection} presents the results of our calculations.
Section \ref{summarysection} closes this work with a discussion and a summary.

\section{Framework and Formalism}\label{frameworksection}

\subsection{Lagrangian density and nonlocality distribution functions}

The generic Euclidean action of the two-flavor PNJL model is
\begin{align}
{\cal S} 
& = \int_0^\beta\mathrm{d}\tau\int \mathrm{d}^3 x 
    \left[ {\bar q}(x) (-i \!\not\!\!D + m_0 ) q(x) 
         - {\cal L}_\mathrm{int} \right]
\nonumber\\
&
             +\beta V\,{\cal U}(\Phi[A],{\bar \Phi}[A];T)\,,
             \label{Lagrangian}
\end{align}
where $q(x)$ is the two-flavor quark field, $m_0$ denotes the current quark mass, and $D^\nu=\partial^\nu + i A^\nu=\partial^\nu + i\delta^{\nu}_{4}\,A^{4,a}{\lambda_a / 2}$ is the color gauge-covariant derivative with $\mathrm{SU}(3)_c$ Gell-Mann matrices $\lambda_a$. 
The gauge coupling $g$ is understood to be absorbed in the definition of
$A^{4,a}$.

The last term in Eq.~(\ref{Lagrangian}) is the Polyakov-loop-effective potential ${\cal U}$, multiplied by volume $V$ and inverse temperature $\beta = T^{-1}$, and to be specified later.
$\Phi$ and ${\bar \Phi}$ are the Polyakov loop and its conjugate, respectively.

The nonlocal generalization of the PNJL model is characterized by an interaction featuring nonlocal quark currents and densities, as follows \cite{Hell:2009,Hell:2010qc,Contrera:2010ik,Hell:2011,Pagura:2011ho}: 
\begin{align}
{\cal L}_\mathrm{int} (x) 
&= G~\Bigl[ j_\mathrm{a} (x) j_\mathrm{a} (x) 
          + J (x) J(x) 
     \Bigr]
\nonumber\\
&- G_\mathrm{v}~j_\mathrm{v}^\mu (x) j_{\mathrm{v},\mu} (x) \,,
\\
j_\mathrm{a} (x) 
&= \int \mathrm{d}^4z~ 
   \tilde{\cal C}(z)~{\bar q}( x + z/2)~\Gamma_\mathrm{a}~q( x-z/2)\,,\label{ja}
\\
j_\mathrm{v}^\mu (x) 
&= \int \mathrm{d}^4z~ 
   \tilde{\cal C}(z)~{\bar q}( x + z/2)~\gamma^\mu~q( x-z/2)\,,\label{jv}
\\
J (x)
&= \int \mathrm{d}^4z~ 
   \tilde{\cal F}(z)~ {\bar q}(x + z/2)~
   \frac{i\delsla^{^{^{\!\!\!\!\!\leftrightarrow}}}}{2 \kappa}~q(x - z/2)\,.\label{capj}
\end{align}
The chiral (scalar and pseudoscalar) densities $j_\mathrm{a}(x)$ with
$\mathrm{a} = 0,1,2,3$ involve the operators
$\Gamma_\mathrm{a}=(1,i\gamma_5 {\vec \tau})$. 
The overall coupling strength $G$ of dimension $[\text{length}]^2$ is chosen sufficiently large so that spontaneous chiral symmetry breaking and pions as Goldstone bosons emerge properly.
The term involving the nonlocal quark vector currents $j_\mathrm{v}^\mu$ has a coupling strength $G_\mathrm{v}$, again of dimension $[{\rm length}]^2$. This $G_\mathrm{v}$ is treated as a parameter in the present work. For orientation, the Fierz transformation of a color-octet current-current interaction (induced, e.\,g., by gluon exchange) gives $G/G_\mathrm{v}=1/2$ (see, e.\,g., Ref.~\cite{Kashiwa:2011}).

The term involving $J(x)$ is an additional vector-type derivative coupling with $~{\bar q}(x')\,\partial^{^{^{\!\!\!\!\!\leftrightarrow}}}_\mu\,q(x) := {\bar q}(x') (\partial_\mu q) (x) - (\partial_\mu {\bar q})(x')\,
q(x)$ together with a scale $\kappa$ so that the effective strength of this term in ${\cal L}_\mathrm{int}$ is $G/\kappa^2$. 
In the following, we refer to the interaction induced by $J(x)$ simply as a \emph{derivative coupling} in order to avoid confusion with the \emph{nonderivative vector interaction} which is called \emph{vector-current interaction} from here on.

The currents Eqs.~\eqref{ja}--\eqref{capj} include nonlocality distributions $\tilde{\cal C}(z)$ and $\tilde{\cal F}(z)$. 
These distributions govern the momentum dependences of the quark mass function and of the renormalization factor that appears in the quark quasi-particle propagator, $Z(p^2)(\gamma\cdot p - M(p^2))^{-1}$,
~\cite{Contrera:2010ik,Hell:2011,Kashiwa:2011}.
The Fourier transform ${\cal C}(p^2)$ of $\tilde{\cal C}(z)$ is related to the quasi-particle mass function $M(p^2)$ determined by the self-consistent gap equation,
\begin{align}
M(p^2) &= Z(p^2) 
\Bigl[ m_0 + \sigma\,{\cal C}(p^2) \Bigr]\,,
\end{align}
where $\sigma$ is the scalar mean field basically representing the chiral condensate 
$\langle\bar{q}q\rangle$.  The Fourier transform ${\cal F}(p^2)$ of $\tilde{\cal F}(z)$ is, in turn, related to  the $Z$ factor of quark wave-function renormalization,
\begin{align}
Z(p^2) &= \Bigl[ 1 - \frac{v}{\kappa}\, {\cal F}(p^2) \Bigr]^{-1}\,,
\end{align} 
where $v$ is the mean field induced by $J(x)$~\cite{Noguera:2008yu}.

The following four-dimensional momentum-space forms of the distribution functions are used in this study:
\begin{align}
{\cal C} (p^2) 
&= \int \mathrm{d}^4z \,\exp(-i p\cdot z)\, \tilde{\cal C}(z)
\nonumber\\
& = \begin{cases}
e^{-p^2 d_C^2/2} & (p^2< \lambda^2)\\
{\cal N}\,\frac{\alpha_s(p^2)}{p^2}&(p^2 \geq \lambda^2)~\,,\end{cases}
\label{eq10}
\\
{\cal F}(p^2) 
&= \int \mathrm{d}^4z \,\exp(-i p\cdot z)\, \tilde{\cal F}(z) 
\nonumber\\
&=  \exp \Bigl( -p^2 d_F^2/2 \Bigr)\,.
\label{eq:form_factor}
\end{align}
The running QCD coupling $\alpha_s(p^2)$ determines the asymptotic form of ${\cal C}(p^2)$ while its infrared behavior is given by a Gaussian parameterization with a characteristic length scale $d_C$.
The matching of these high- and low-momentum representations at an intermediate scale $\lambda$ determines the constant ${\cal N}$.
Parameters in both distribution functions are fitted to LQCD data as described in Ref.~\cite{Hell:2011}.

\subsection{Thermodynamical potential}

Consider now the (grand-canonical) thermodynamical potential, 
	\begin{equation}
		\Omega=-\dfrac{T}{V}\ln\mathcal{Z}\,,
		\end{equation}
where 
	\begin{equation}
	\mathcal{Z}=\int\mathcal{D}q\,\mathcal{D}\bar q\,{\rm e}^{-\mathcal{S}}
	\end{equation}
is the grand-canonical partition function determined by the path integral over the action \eqref{Lagrangian}.
In the mean-field approximation the fields are replaced by their (thermal) expectation values. After bosonization, the mean-field  thermodynamical potential, $\Omega_{\rm MF}$ of the nonlocal PNJL model, including quark wave-function-renormalization corrections, but in the absence of the the vector-current interaction reads
\begin{align}
\Omega_{\rm MF} &= \Omega_1 + {\cal U}(\Phi,{\bar \Phi};T)\,,
\label{TP-PNJL}
\end{align}
where
\begin{align}
\Omega_1 
&= - 4T \sum_{i=\pm,0} \sum_{n=-\infty}^{\infty}
           \int \frac{\mathrm{d}^3 p}{(2\pi)^3} 
\ln \left[ \dfrac{\omega_{n,i}^2 + E_i^2(p^2)}{Z_i^2(p^2)} \right]
\nonumber\\
&+  \frac{ \sigma^2 + v^2}{4G}.
\label{Eq:TP}
\end{align}
Here $\sigma$ and $v$ are the mean fields associated with the scalar density $j_0$ and the derivative vector current $J$, respectively. 
The first term on the right-hand side of Eq.~(\ref{Eq:TP})
involves the quark quasi-particle energies
\begin{align}
E_i &= \sqrt{\vec{p}\,^2+M_i^2(p^2)}
\end{align}
with dynamically generated masses, $M_i(p^2) \equiv M (p^2 =
\omega_{n,i}^2 + \vec{p}\,^2)$, determined self-consistently at each
shifted Matsubara frequency $\omega_{n,i}$ with $i \in \{ \pm,0\}$:
\begin{align}
\omega_{n,\pm} &= \omega_n  - i \mu \pm \frac{A_4^3}{2} -
 \frac{A_4^8}{2\sqrt{3}}~~,\nonumber \\
\omega_{n,0  } &= \omega_n - i \mu + \frac{A_4^8}{\sqrt{3}}\,.
\label{eq17}
\end{align}
$A_4^{3,8}$ are the gauge fields forming the Polyakov loop given
in Eq.~\eqref{eq:polyakovloop}.
Likewise, the $Z$ factors are understood as $Z_i(p^2)\equiv Z(p^2=\omega_{n,i}^2+\vec{p}\,^2)$.
More explicitly:
\begin{align}
M_i(p^2) 
&= Z_i(p^2) \Bigl[ m_0 + \sigma\,{\cal C}(p^2=\omega_{n,i}^2+\vec{p}\,^2) \Bigr]\,,\\
Z_i(p^2)    &= \Bigl[ 1 - \frac{v}{\kappa}\, 
{\cal F}(p^2 =\omega_{n,i}^2+\vec{p}\,^2) \Bigr]^{-1}\,.
\end{align}

At finite temperature $T$, the Lorentz invariance is broken by the thermal medium and the inverse quark quasi-particle propagator becomes
$S^{-1}(p) =- {\cal A}_4(p)\, \gamma_4 p^4- {\cal A}(p) \gamma_i p^i +
{\cal B}(p)$ with ${\cal A}_4 \neq {\cal A}$. 
Here we assume for simplicity that the difference  between ${\cal A}_4$ and ${\cal A}$ is sufficiently  small so that it can be neglected, given that the overall influence of wave-function renormalization on thermodynamical quantities is not very significant.

The introduction of the vector-current interaction leads to the following modifications of the thermodynamical potential \eqref{TP-PNJL}: first, from the bosonization of Eq.~\eqref{jv} a quadratic term involving the vector mean field $\omega$,
\begin{align}
- \frac{ \omega^2}{4G_\mathrm{v}}\,,
\end{align}
is added to the thermodynamical potential (\ref{Eq:TP}); second, 
 the chemical potential is shifted according to 
$\mu \to \mu 
- {\cal C}(p^2=\omega_{n,i}^2 + {\vec p\,}^2)\ \omega$.
 The vector mean field basically represents the baryon density, $\varrho=\langle j_{\rm v}^0\rangle$, in the form $\omega=\frac{\varrho}{2G_\mathrm{v}}$.
One important remark is in order: the $\omega$-dependence does not appear in the distribution functions because these functions are introduced in the Lagrangian density before taking the mean-field approximation.

\subsection{Polyakov-loop potential}
In this study, we consider two types of the Polyakov-loop effective potentials.
The first one is given in Ref.~\cite{Ratti:2006wg}: 
\begin{align}
&{{\cal U} ({\bar \Phi},\Phi;T)\over T^4}
= -\frac{1}{2}\, b_2(T)\, {\bar \Phi}\, \Phi
\nonumber\\
&
              + b_4(T)\, \ln[ 1 - 6\, {\bar \Phi} \,\Phi 
                            +4 ({\bar \Phi}^3 + \Phi^3) 
                             -3 ({\bar \Phi}\,\Phi)^2]\,, 
\label{PLPa}
\end{align}
where $\Phi$ and ${\bar \Phi}$ are represented as
\begin{align}
\Phi 
&= \frac{1}{3} 
   \Bigl[ \exp\left(i {A_4^3 + A_4^8\over 2T}\right) 
        + \exp\left(-i {A_4^3 - A_4^8\over 2T}\right)  
\nonumber\\
&       + \exp\left(i {A_4^8\over \sqrt{3}T}\right) \Bigr]\,,
\label{eq:polyakovloop}\\
{\bar \Phi} &= \Phi^*.
\end{align}
The other one is proposed in Ref.~\cite{Fukushima:2004}:
\begin{align}
\frac{{\cal U} ({\bar \Phi},\Phi;T) }{T^4}
&= - bT \Bigl[ 54 e^{-\frac{a}{T}} {\bar \Phi}\, \Phi
\nonumber\\
& \hspace{-1.5cm}
             + b_4(T)\, \ln\{  1 - 6\, {\bar \Phi} \,\Phi 
                             + 4 ({\bar \Phi}^3 + \Phi^3) 
                             - 3 ({\bar \Phi}\,\Phi)^2 \}
\Bigr]\,.
\label{PLPb}
\end{align}
This latter form is obtained from the knowledge of the strong-coupling limit of
QCD. Recently the details have been investigated in Ref.~\cite{Sasaki:2012,Ruggieri:2012}.
The nonlocal PNJL model with potential (\ref{PLPa}) is henceforth denoted as model A and
that with potential (\ref{PLPb}) as model B.

It is convenient to introduce a modified Polyakov-loop
and its conjugate as
\begin{align}
\Psi &= e^{i\theta} \Phi\,,~~~~{\bar \Psi} = e^{-i\theta} {\bar \Phi}\,,
\end{align}
as these are RW-periodic quantities.
The real and imaginary parts of $\Psi$ serve as order parameters of the deconfinement transition and spontaneous $\mathcal{C}$-symmetry breaking because $\mathrm{Im}\,\Psi$ is a $\theta$-odd quantity, just like the quark number density.
We use $\mathrm{Im}\,\Psi$ as the
order parameter of  ${\cal C}$-symmetry breaking. As mentioned in the introduction, this $\theta$-odd quantity serves as an exact
order parameter at $\theta=\pi/3$ because there $\mathcal{C}$-symmetry is not explicitly broken.

\subsection{Parameter setting}

In the PNJL model the pion mass and its decay constant are used to fix parameters in the NJL sector of the Lagrangian. 
These parameters are taken from Ref.~\cite{Hell:2011}. The vector-current interaction, at mean-field level, has no influence on the thermodynamics at $\mu=0$. An estimate of the coupling constant $G_{\rm v}$ can therefore only be provided by comparison with (restricted) LQCD information at nonzero chemical potential. For guidance, we can use the LQCD value for the ratio $T_{\rm RW}/T_{\rm c}\sim1.05$ \cite{Forcrand:2002ls}, where $T_{\rm RW}$ is the critical temperature of the Roberge-Weiss end point (at $\theta=\pi/3$) and $T_{\rm c}$ is the crossover temperature at $\theta=0$.

The coefficient functions $b_2(T)$ and $b_4(T)$ in (\ref{PLPb}) are parameterized such as to reproduce pure-gauge-LQCD results (Refs.~\cite{Hell:2010qc,Hell:2011}). 
When we use $T_0=270\,{\rm MeV}$ for the confinement-deconfinement transition temperature in the pure-gauge case, the resulting crossover transition temperature when including quarks in the PNJL model is slightly higher than the LQCD prediction.
Alternatively, we also use $T_0=240\,{\rm MeV}$ in order to reproduce $T_\mathrm{c} \sim 190\,{\rm{MeV}}$, as suggested in Ref.~\cite{Schaefer:2007pw}.

The parameter $a$ in (\ref{PLPb}) is fitted to reproduce the critical temperature in the pure gauge limit and its  value is $a=664\,{\rm MeV}$.
The remaining parameter $b$ is defined to reproduce the pseudo-critical temperature with dynamical quarks.
To reproduce $T_\mathrm{c} \sim 190\,{\rm MeV}$ from the two-flavor LQCD data, we take $b=0.01$.

\section{Numerical results}\label{resultssection}

\begin{figure}[htbp]
\begin{center}
 \includegraphics[width=0.23\textwidth]{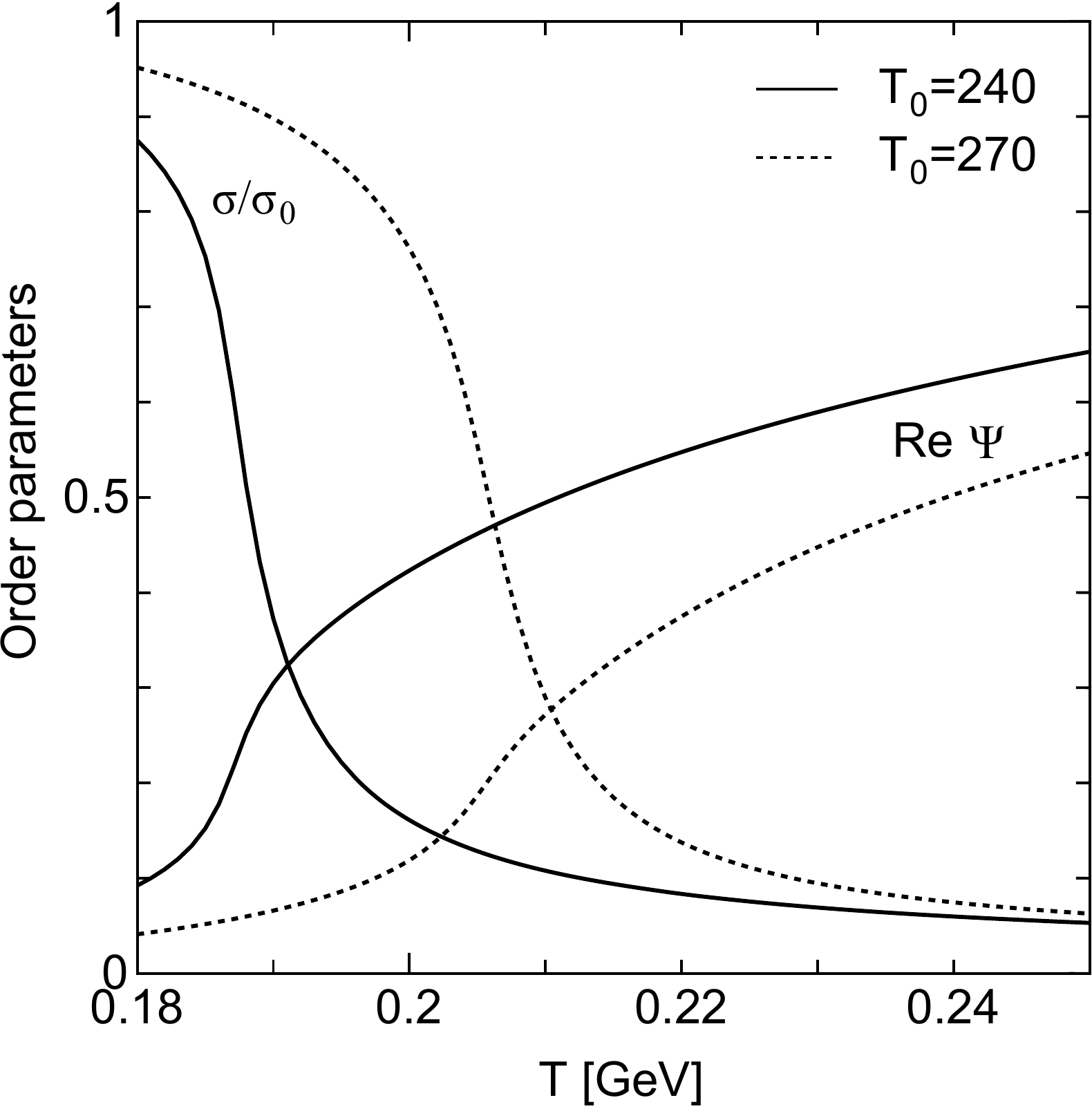}
 \includegraphics[width=0.23\textwidth]{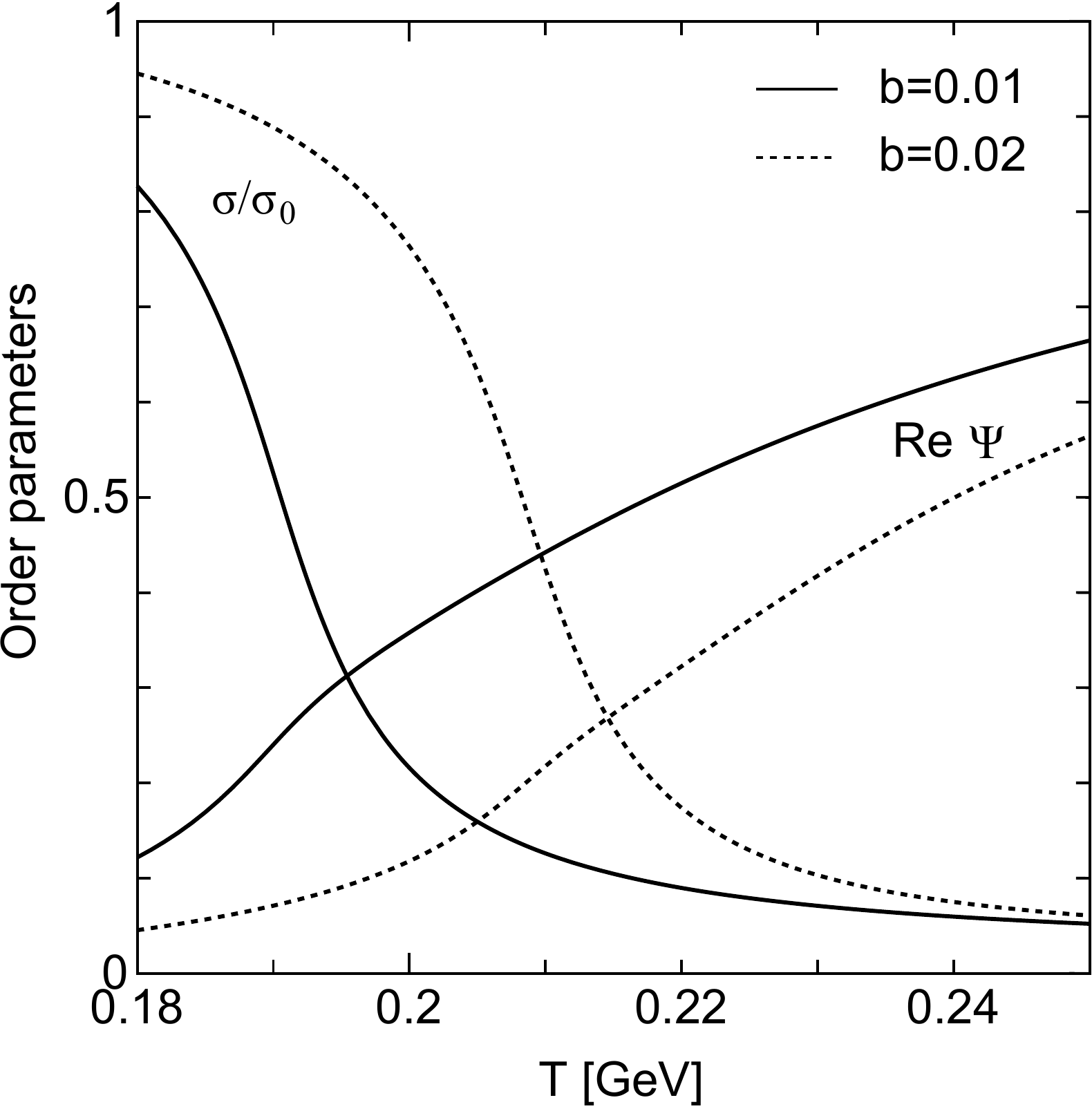}
\end{center}
\vspace{-.5cm}
\caption{ $T$-dependence of the chiral order parameter  and the real part of $\Psi$ at $\mu=0$ with $G_\mathrm{v}=0$. 
The solid and dotted lines denote the result of  model A with
$T_0=240$ and $270\,\rm{ MeV}$, respectively.
The left figure shows the result of  model B with $b=0.01$ and $0.02$.
}
\label{Fig:b-dep}
\end{figure}

Figure~\ref{Fig:b-dep} displays the $T$-dependence of the chiral order parameter and the real part of $\Psi$.
Here we show the results of  model A with $T_0=240$ and $270\,{\rm MeV}$ and that of  model B with $b=0.01$ and $0.02$ at $\mu=0$, in both cases with
$G_\mathrm{v}=0$.
In all cases the  transitions are crossovers. The transition temperatures for the chiral and deconfinement crossovers almost coincide.
This property comes from the entanglement of the chiral and deconfinement transitions through the nonlocality distribution functions.

\begin{figure}[htbp]
\begin{center}
 \includegraphics[width=0.23\textwidth]{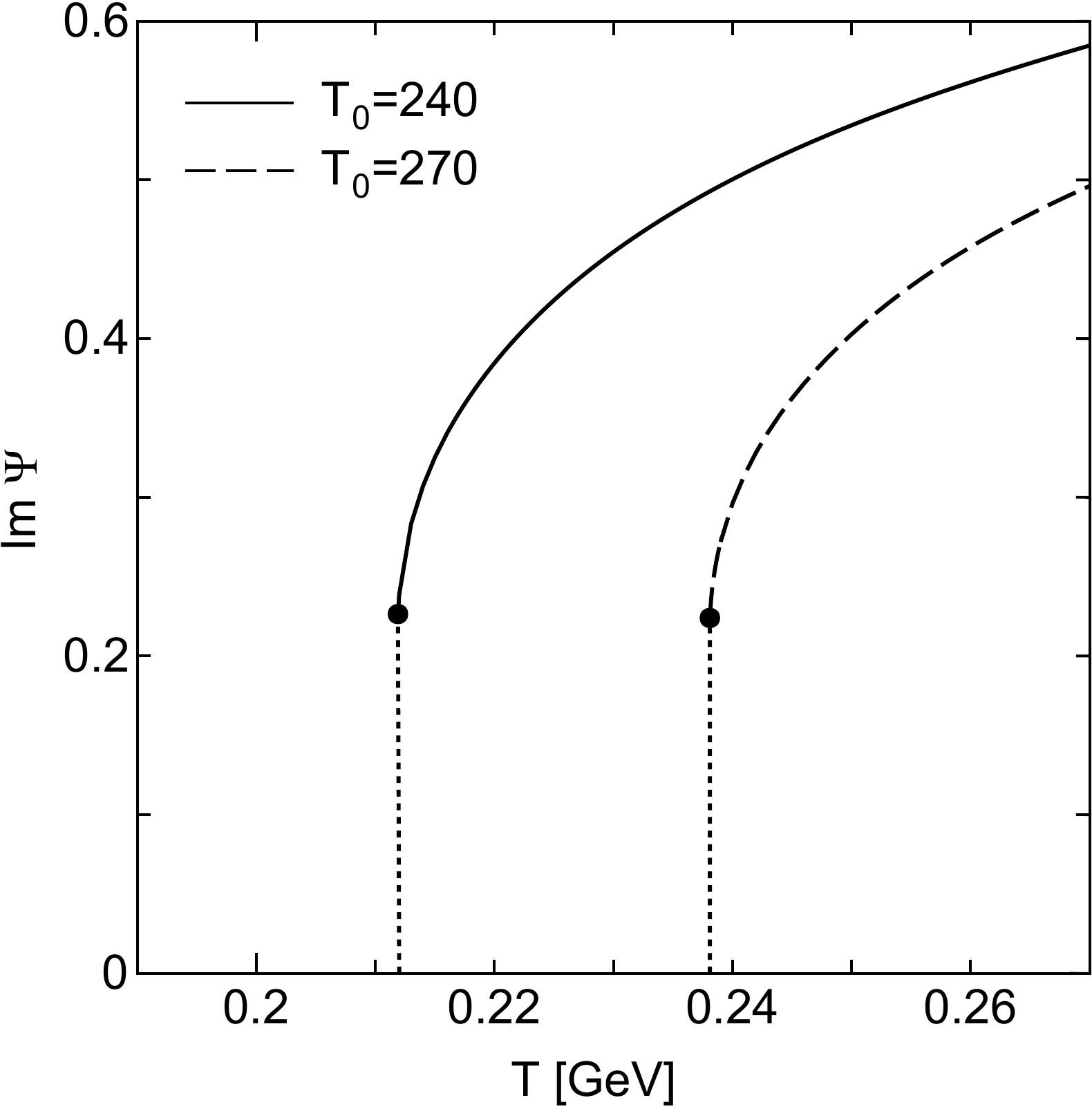}
 \includegraphics[width=0.23\textwidth]{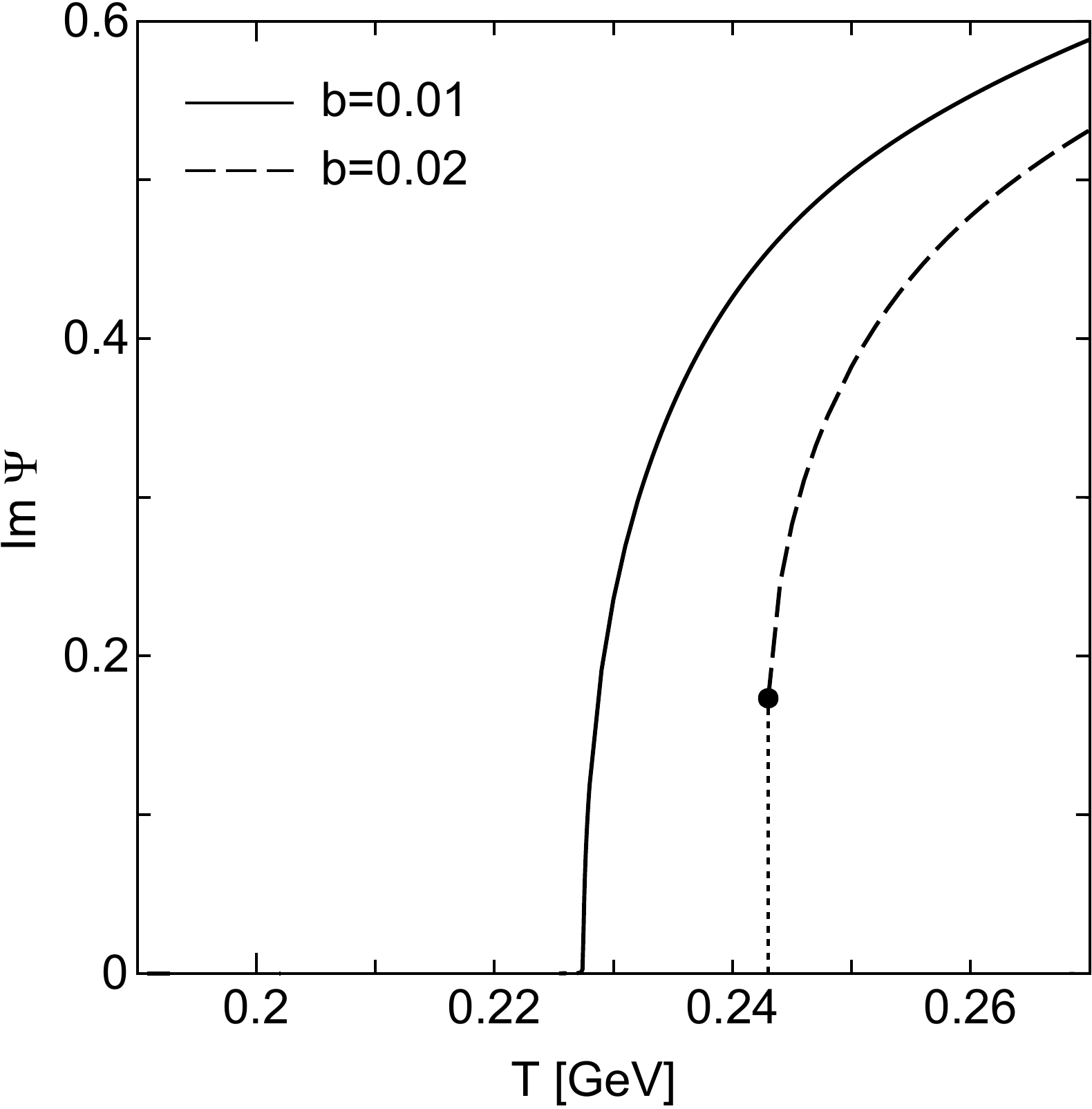}
\end{center}
\vspace{-.5cm}
\caption{ Left (right) figure: $T_0$-dependence
($b$-dependence) of $\mathrm{Im}~\Psi$ as function of temperature $T$ at $\theta=\pi/3$ with
$G_\mathrm{v}=0$. 
In the left figure, the solid and dotted lines denote the result of model A with $T_0=240$ and $270$ MeV, respectively.
In the right figure, the solid and dotted lines denote the result of  model B with $b=0.01$ and $0.02$, respectively.
}
\label{Fig:T-b-dep}
\end{figure}

Figure~\ref{Fig:T-b-dep} shows the $T$-dependence of $\mathrm{Im}\,\Psi$  at $\theta=\pi/3$.
From the right figure we  see that there is a first-order RW end point in the case of $b=0.02$ which turns into second-order for $b=0.01$. 
The LQCD data~\cite{Bonati:2011a,Bonati:2011b} suggest that the order of the RW end point is first-order at sufficiently small $m_0$.
From this perspective, $b=0.01$ is not a suitable choice, but this situation can be modified as shown below.

\begin{figure}[htbp]
\begin{center}
 \includegraphics[width=0.23\textwidth]{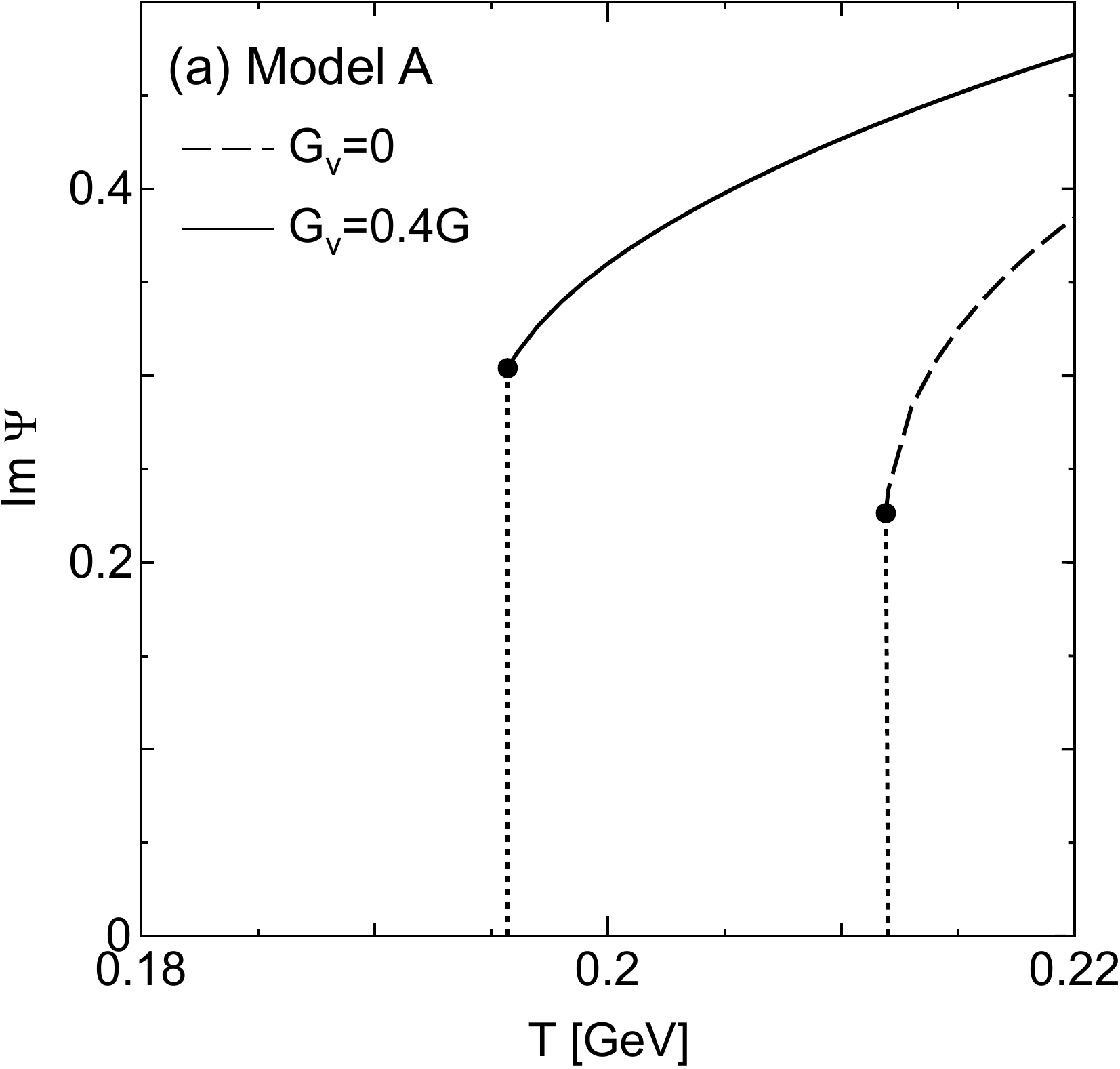}
 \includegraphics[width=0.23\textwidth]{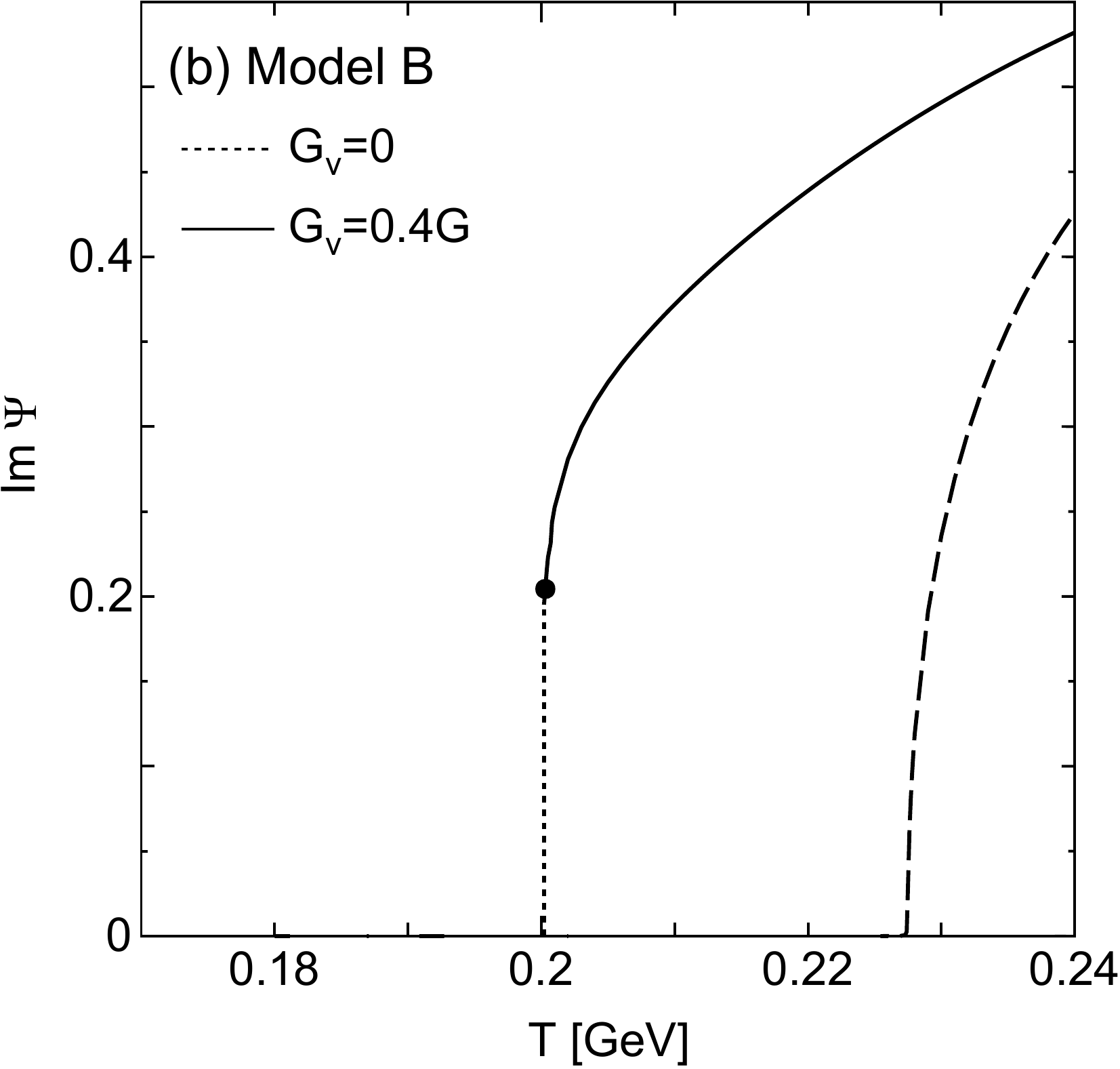}
\end{center}
\vspace{-.5cm}
\caption{  $T$-dependence of $\mathrm{Im}~\Psi$ at
$\theta=\pi/3$ for $G_\mathrm{v}=0$ and $0.4\, G$. 
The solid and dotted lines denote the result with $G_\mathrm{v}=0.4\, G$ and $G_\mathrm{v}=0$, respectively.  Left figure:  result of  model A with $T_0=240\,{\rm MeV}$; rigth figure: model B with $b=0.01$.
}
\label{Fig:T-b-Gv-dep}
\end{figure}

In the previous figures we have ignored the vector-current interaction. As a consequence, the ratio $T_\mathrm{RW}/T_\mathrm{c}$ exceeds the LQCD prediction~\cite{Wu:2007iu,Forcrand:2002ls}.
Choosing $G_\mathrm{v}=0.4\, G$ in model A with $T_0=240\,{\rm MeV}$, this ratio becomes  $T_\mathrm{RW}/T_\mathrm{c}=1.04$ \cite{Kashiwa:2011}.
Model B with $G_\mathrm{v}=0.4\, G$ and $b=0.01$ leads to $T_\mathrm{RW}/T_\mathrm{c} \simeq 1.05$.
These values are close to the LQCD result.
Figure~\ref{Fig:T-b-Gv-dep} shows again the $T$-dependence of the imaginary part of the modified Polyakov loop ($\mathrm{Im}~\Psi$).
One observes that the transition at $T_\mathrm{RW}$ becomes  first-order when introducing the vector-current interaction in  model B with
$b=0.01$.
Henceforth, we only refer to the results of  model A as both models lead to almost identical results.

\begin{figure}[htbp]
\begin{center}
 \includegraphics[width=0.40\textwidth]{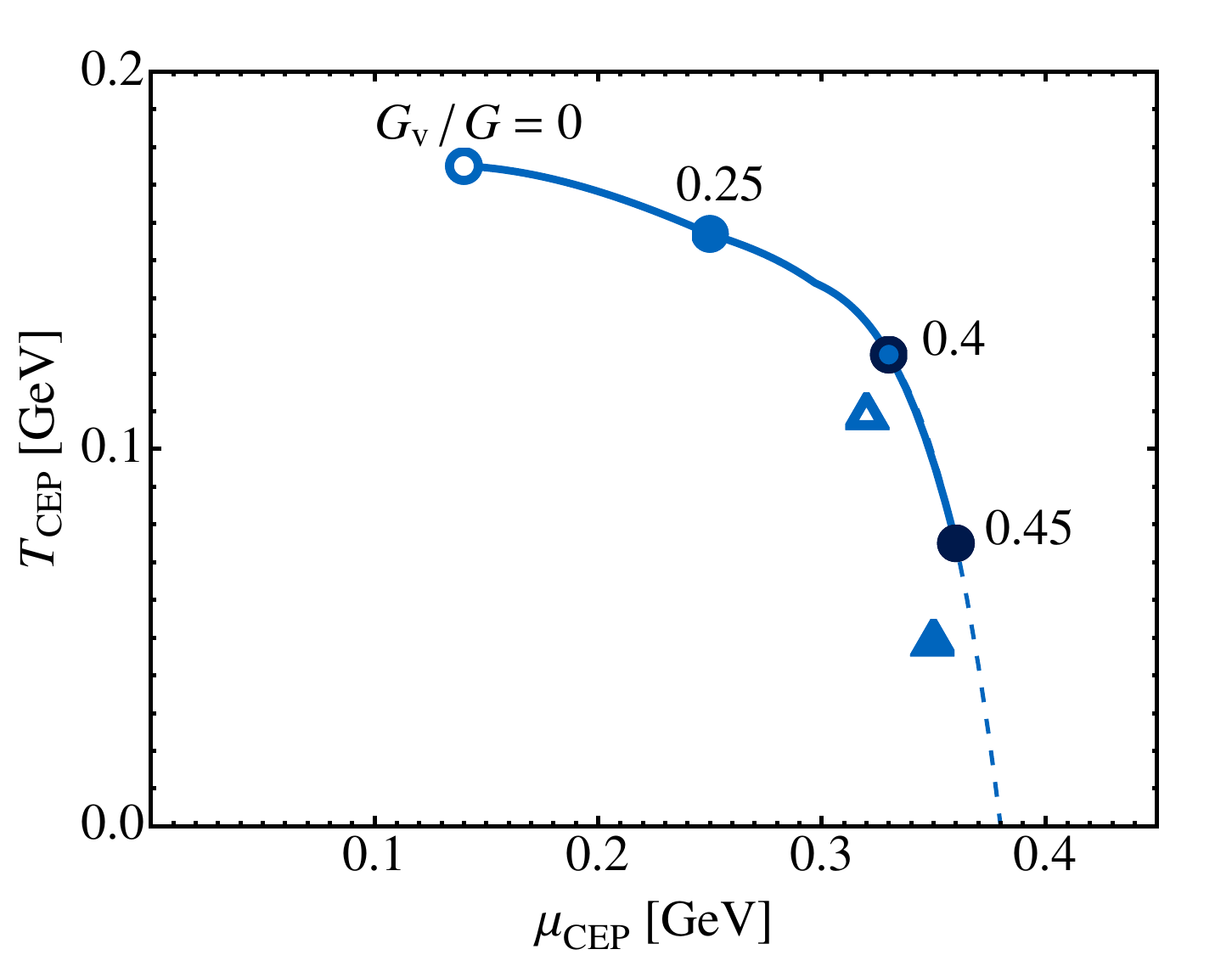}
\end{center}
\vspace{-.5cm}
\caption{ Phase diagram in the $\mu_\mathrm{R}$-$T$  plane. The circles and triangles represent the position of the critical end point for the nonlocal and the local PNJL models, respectively.
The white, blue, black blue and black symbols correspond to results with
$G_\mathrm{v}/G=0$, $0.25$, $0.4$ and $0.45$, respectively.
}
\label{Fig:PD}
\end{figure}

Finally, we study the $G_\mathrm{v}$-dependence of the position of these critical point in the $T$-$\mu$ phase diagram.
In Fig.~\ref{Fig:PD}, the circles and triangles represent the positions of the critical end points for the nonlocal version of the PNJL model, respectively. Results are presented for
$G_\mathrm{v}/G=0$, $0.25$, $0.4$ and $0.45$, respectively. 
In the local PNJL model, the critical end point disappears or shifts to very small $T$ when considering a realistic range  $0.25\lesssim G_\mathrm{v}/G\lesssim 1$ for the vector coupling strength.
In the nonlocal PNJL model, the location of the critical point has a less pronounced dependence on temperature, at least for small $G_\mathrm{v}$. This behavior can be traced to the weakening of the NJL interaction by the nonlocality distribution. The downward trajectory of the critical point becomes very steep, however, once $G_\mathrm{v}/G$ reaches values of $0.4$ and beyond (see Fig.~\ref{Fig:PD}). Around $G_\mathrm{v}/G\simeq 0.5$, the canonical ratio corresponding to an effective quark-quark interaction induced by color-octet (gluon-exchange) currents, the critical points tends to disappear altogether and the first-order phase transition turns into a continuous crossover.

\section{Summary}\label{summarysection}

In this study we have investigated the impact of a (nonderivative) vector-current interaction in the nonlocal PNJL model at real and imaginary chemical potentials.
The presence of the vector-current interaction makes the transition at the Roberge-Weiss end point more pronounced. The RW end point becomes first-order, consistent with  recent LQCD simulations.

The location of the critical point in the phase diagram for real chemical potentials is highly sensitive to the vector coupling strength $G_{\rm v}$. In the nonlocal PNJL model used here, the critical point tends to be eliminated in favor of a continuous crossover once the ratio of vector-to-scalar couplings is increased toward and beyond $G_{\rm v}/G=1/2$, the value characteristic of an effective gluon-exchange interaction between quarks. Qualitatively similar tendencies are found in recent related work \cite{Contrera:2012wj} and in a $2+1$-flavor study using the local PNJL model \cite{Bratovic:2012qs} in which the disappearance of the chiral first-order transition turning to a crossover is indicated already at values $G_{\rm v}/G<1/2$.

\noindent
\begin{acknowledgments}
K.~K.\ thanks  H.~Kouno and M.~Yahiro for fruitful discussions.
K.~K.\ is supported by RIKEN Special Postdoctoral Researchers Program. T.~H.\ acknowledges the kind hospitality at Brookhaven National Laboratory during his stay.
This work is supported in part by BMBF,  by the Excellence Cluster ``Origin and
Structure of the Universe'', and by DFG and NSFC through CRC 110.
\end{acknowledgments}

\end{document}